\documentclass[twoside,reqno]{HERON}
\usepackage{epsfig,cite,colordvi}
\usepackage{graphicx}
\usepackage{url}
\usepackage{amssymb,amsmath,amscd,epsf}
\usepackage{times}
\usepackage{makeidx}
\newcommand{\bea}{\begin{eqnarray}}
\newcommand{\eea}{\end{eqnarray}}
\newcommand{\be}{\begin{equation}}
\newcommand{\ee}{\end{equation}}

\newcommand{\nr}{{\bf r}}

\newcommand{\nq}{{\bf q}}
\newcommand{\np}{{\bf p}} 
\newcommand{\nP}{{\bf P}}

\begin{document}

\title{Short-Range Correlations and Meson-Exchage Currents in Electron
  and Neutrino Scattering}
\runningheads{Short-Range Correlations and Meson-Exchage Currents}
{P.R.Casale and J.E. Amaro}

\begin{start}

  \author{P.R. Casale }{1,2} and \coauthor{J.E. Amaro}{1,2}
  
\index{Casale, P.R.}
\index{Amaro, J.E.}

\address{Departamento de Fisica Atomica, Molecular y Nuclear, 
Universidad de Granada, 18071 Granada, Spain}{1}
\address{Instituto Carlos I de Fisica Teorica y Computacional,
Universidad de Granada, 18071 Granada, Spain}{2}

\begin{Abstract}
  We investigate meson-exchange currents (MEC) in the one-particle
emission transverse response of nuclear matter, incorporating
short-range correlations via the Bethe-Goldstone equation with a
realistic nucleon-nucleon interaction. The interference between
one-body and two-body currents, strengthened by the high-momentum
components of correlated pairs, produces a marked enhancement of the
transverse response. We also indicate how the formalism extends to
neutrino scattering, where similar effects are expected to impact
oscillation experiments.

\end{Abstract}
\end{start}

\section{Introduction}
There is clear experimental evidence of an enhancement of the
transverse response in the quasielastic peak with respect to
single-particle models in the 1p1h channel \cite{Jou96,Bod22}.
Two-particle–two-hole (2p2h) excitations induced by meson-exchange
currents (MEC) are not sufficient to explain this effect, since their
main contribution lies in the dip region between quasielastic and pion
production \cite{Ama94b}.  Ab initio calculations in light nuclei also
find a transverse enhancement produced by MEC, although they cannot
disentangle between 1p1h and 2p2h channels \cite{Car02,Lov16}.  For
heavier nuclei and nuclear matter, however, single-particle models
consistently predict that MEC reduce the 1p1h transverse response,
mainly due to the $\Delta$ current
\cite{Koh81,Alb90,Ama94a,Ama03,Cas23b}, a result corroborated in
recent comparisons across different single-particle approaches,
including the spectral function model \cite{Cas25}.  In contrast, the
calculation of Fabrocini within the correlated basis function (CBF)
framework found an enhancement in nuclear matter due to the combined
effect of short-range correlations (SRC) and MEC \cite{Fab97}.

Motivated by this, in the present work we develop a correlated model
of the transverse response in the 1p1h channel that includes MEC
within the independent pair approximation.  This is made possible by
our recent solution of the Bethe-Goldstone equation for two correlated
nucleons in nuclear matter, which provides wave functions with
explicit high-momentum components \cite{Cas23}.  Here we apply this
framework to investigate whether correlations together with MEC are
sufficient to account for the observed transverse enhancement.

\section{Analysis of SRC using the Bethe-Goldstone equation}

In order to describe short-range correlations (SRC) in nuclear matter
we solve the Bethe-Goldstone (BG) equation for a correlated pair of
nucleons \cite{Bet57,Gol57}.
We assume that two nucleons interact in the presence of a
degenerate Fermi gas. The Pauli exclusion principle forbids them from
scattering into already occupied states, and as a consequence their
wave function is modified by the interaction and acquires
high-momentum components above the Fermi surface,
\begin{equation}
 \Psi = \Psi_{0} + Q \Psi ,
\end{equation}
where $Q$ is the Pauli blocking projector over high momenta $>k_F$,
\begin{equation}
Q  |p'_1 p'_2 \rangle=\left\{
\begin{array}{lcl}
  |p'_1 p'_2 \rangle
  & \text{if} & |p'_i| > k_F, \\
0 & & \text{otherwise.}
\end{array}
\right.
\end{equation}
We assume that the uncorrelated and correlated wave functions satisfy the two-body Schrodinger equation with the same energy
\begin{equation}
T \Psi_{0} = E \Psi_{0}, \kern 2cm
(T+V)\Psi = E \Psi ,
\end{equation}
where $V$ is the NN potential.
Operating with $(E-T)$ and solving for $Q\Psi$, one obtains
the BG equation for a correlated pair,
\begin{equation}
\Psi = \Psi_{0} + \frac{Q}{E-T} V \Psi .
\end{equation}
Expressed in integral form, the BG equation reads
\begin{equation}
|\Psi \rangle = |\nP,\np \rangle 
 + \int d^{3}\nP' d^{3}\np' 
\frac{Q(\nP',\np')}{\tfrac{(\nP^{2}-\nP'^{2})}{2M_T}-\tfrac{(\np^{2}-\np'^{2})}{2\mu}} 
|\nP',\np'\rangle 
\langle \nP',\np'|V|\Psi\rangle . \nonumber
\end{equation}
where $\nP=\np_1+\np_2$ is the total momentum of two nucleons in the
Fermi gas, below the Fermi momentum, $p_1,p_2 < k_F$, and
$\np=(\np_1-\np_2)/2$ is the relative momentum.  Since the
nucleon-nucleon potential depends only on the relative coordinate
$\nr=\nr_1-\nr_2$,
\begin{equation}
\langle \nP',\np' | V | \Psi \rangle =
\delta(\nP'-\nP)\,\langle \np' | V | \psi \rangle,
\end{equation}
where $\psi(\nr)$ is the relative wave function of the nucleon pair, verying
the integral BG equation
\begin{equation}
|\psi \rangle = |\np \rangle + \int d^{3}\np'\;
\frac{Q(\nP,\np')}{p^{2}-p'^{2}} |\np'\rangle
\langle \np'| 2\mu V |\Psi \rangle .
\end{equation}
This equation is further simplified by performing an angular average
of the Pauli blocking operator $Q$, so that it depends only on the
moduli of the center-of-mass momentum $P$ and the relative momentum
$p$ \cite{Cas23}.  This simplification is convenient for performing
the multipole expansion of the wave function.

The correlated relative wave function is a bispinor, reflecting the
fact that the nucleons are spin-$\frac{1}{2}$ particles. Since the
nucleon-nucleon potential does not mix spin channels, we can treat
separately the cases singlet ($S=0$) and triplet ($S=1$) and perform a
multipole expansion,
\begin{equation}
|\psi \rangle = |\np,SM_s \rangle + |\Delta \psi \rangle ,
\end{equation}
\begin{equation}
\langle \nr|\np,S,M_s \rangle =
\sqrt{\frac{2}{\pi}}
\sum_{JM}\sum_{lm_l}
i^l Y^{*}_{lm_l}(\hat{\np}) \langle lm_l S M_s | JM \rangle
j_l(pr) \,\mathcal{Y}_{lSJM}(\hat{\nr}),
\end{equation}
\begin{equation}
\langle\nr|\Delta \psi \rangle =
\sqrt{\frac{2}{\pi}}
\sum_{JM}\sum_{l l' m_l}
i^{l'} Y^{*}_{l'm_l}(\hat{\np}) \langle l'm_l S M_s|JM\rangle
\Delta \phi^{SJ}_{ll'}(r)
\,\mathcal{Y}_{lSJM}(\hat{\nr}),
\end{equation}
where 
\begin{equation}
\mathcal{Y}_{lSJM}(\hat{\nr})=
\sum_{m\mu}\langle l m S \mu|JM \rangle \,Y_{lm}(\hat{\nr})\,|S\mu \rangle,
\end{equation}
The functions $\Delta \phi_{ll'}^{SJ}(r)$ represent the high-momentum
components of the radial part of the correlated wave function. By
substituting the partial-wave expansion into the integral
BG equation, one obtains a set of coupled equations for
each spin-total angular momentum channel $SJ$,
\begin{equation}
u^{SJ}_{ll'}(r) = \delta_{ll'} \hat{j}_l(pr)
+ \int dr' \, \hat{G}_l(r,r') \sum_{l_1} U^{SJ}_{l_1 l}(r') \, u^{SJ}_{l_1 l'}(r').
\end{equation}
Here, $u^{SJ}_{ll'}(r)$ are the reduced radial wave functions, defined
as the full radial wave functions multiplied by $r$, while
$\hat{j}_l(pr)$ are the reduced spherical Bessel functions. The kernel
$\hat{G}_l(r,r')$ is the reduced Green function of the equation, which
can be found in detail in Ref.~\cite{Cas23}.  The functions
$U^{SJ}_{l_1 l}(r)$ are the multipoles of the nucleon-nucleon
potential in the  $S,J$ channel, multiplied by the nucleon
mass.

Table 1 summarizes the
partial-wave decomposition of the correlated two-nucleon system,
listing the total spin $S$, total angular momentum $J$, orbital
angular momenta $(l,l')$, and the corresponding channels, including
coupled and uncoupled cases where tensor mixing occurs.

\begin{table}[h!]
\centering
\begin{tabular}{c c c c}
\hline
$S$ & $J$ & $(l,l')$ & Channels \\
\hline
0 & $l=l'$ & any & ${}^{1}S_{0}, {}^{1}P_{1}, {}^{1}D_{2}, {}^{1}F_{3}$ \\
\hline
1 & 0 & $l=l'=1$ & ${}^{3}P_{0}$ \\
1 & $\ge 1$ uncoupled & $l=l'=J$ & ${}^{3}P_{1}, {}^{3}D_{2}, \dots$ \\
1 & 1 coupled & (0,2) & ${}^{3}S_{1}, {}^{3}D_{1}$ with mixing ${}^{3}S_{1}/{}^{3}D_{1}$ \\
1 & 2 coupled & (1,3) & ${}^{3}P_{2}, {}^{3}F_{2}$ with mixing ${}^{3}P_{2}/{}^{3}F_{2}$ \\
1 & 3 coupled & (2,4) & ${}^{3}D_{3}$ \\
\hline
\end{tabular}
\caption{Partial-wave decomposition of the correlated two-nucleon system.}
\end{table}

\subsection{Solution of the BG equation with a delta-shell potential}

We solve the BG equation for the Granada 2013 potential \cite{Nav13},
\begin{equation}
U^{SJ}_{ll'}(r) = \sum_{i=1}^{N} (\lambda_i)^{SJ}_{ll'} \, \delta(r-r_i),
\end{equation}
where there are $N=5$ delta shells
at positions $r_i$ and strengths $(\lambda_i)^{SJ}_{ll'}$.
The potential was obtained through a partial wave
analysis of $np$ and $pp$ scattering data below the pion production
threshold. Its coarse-grained form reflects the fact
that the nucleon wavelength cannot resolve details of the interaction
at distances smaller than $\Delta r \sim 0.6$~fm.
In this case the radial BG equation becomes
\begin{equation}
u^{SJ}_{ll'}(r) = \hat{j}_l(pr)\delta_{ll'} +
\sum_i \hat{G}_l(r,r_i) \sum_{l_1} (\lambda_i)^{SJ}_{l_1 l}\,
u^{SJ}_{l_1 l'}(r_i).
\end{equation}
Taking $r=r_j$ leads to a closed linear system for
$u^{SJ}_{ll'}(r_j)$, which is solved separately for uncoupled and
coupled partial waves,
\begin{equation}
u_{ll'}^{SJ}(r_j) \;=\; \hat{j}_{l}(pr_j)\,\delta_{ll'} 
+ \sum_{i}\hat{G}_{l}(r_{j},r_{i}) 
\sum_{l_{1}} \bigl(\lambda_{i}\bigr)^{SJ}_{l_{1}l}\,
u_{l_{1}l'}^{SJ}(r_{i}) ,
\label{eq:bg-radial}
\end{equation}
where $r_j$ are the discretized mesh points,
The
uncoupled channels lead to a system of 5 linear equations, which can
be solved trivially by matrix inversion, essentially in analytical
form.  The coupled channels correspond to two linear systems of 10
equations each, which are solved in an equally straightforward manner.

\begin{figure}[htb]
   \centering
   \includegraphics[width=8cm,bb=5 420 500 770]{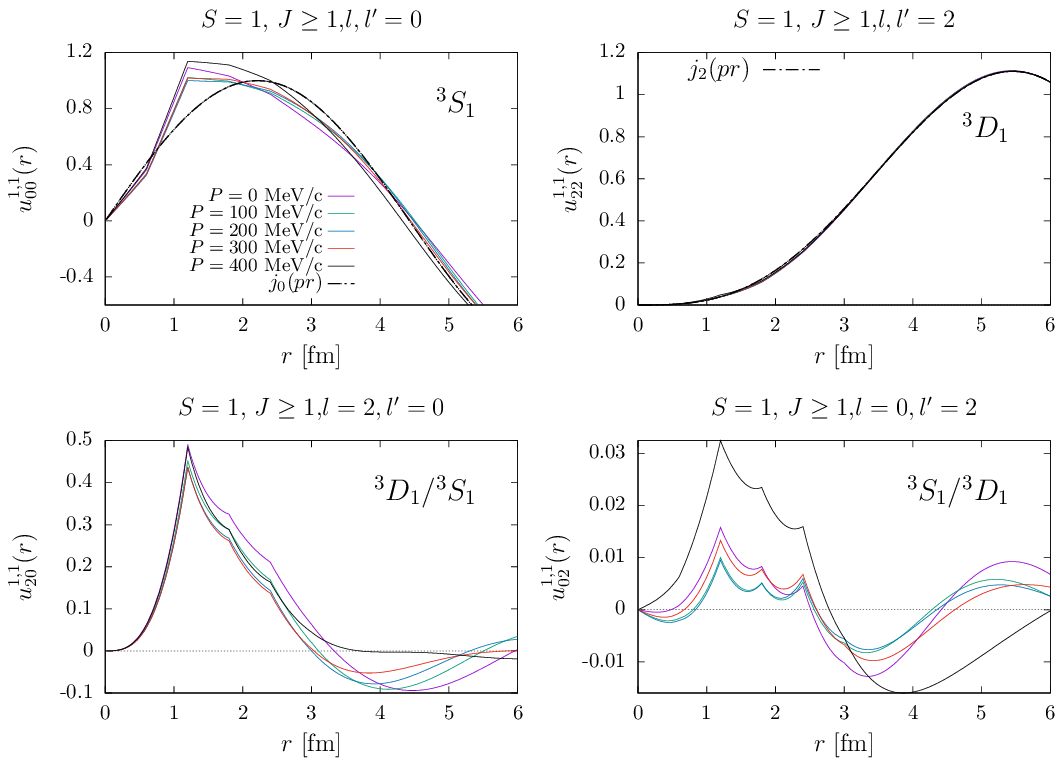}
   \caption{Example of coupled radial wave functions $u_{ll'}^{SJ}(r)$
     obtained from the Bethe--Goldstone equation. The relative
     momentum of the pair is $p=225$ MeV/c. We see the results for
     several values of the total momentum $P$.}
   \label{fig:radialwf}
\end{figure}

In Fig.~1 we show an example of the solutions for the coupled $J=1$
channel.  The solutions correspond to the linear system for
$(l=0,2;\,l'=0)$ and another for $(l=0,2;\,l'=2)$.  The most important
contribution arises from the $(l,l' = 0,2;\, l'=0)$ case. 
In particular, we observe a strong mixing between the ${}^{3}S_{1}$
and ${}^{3}D_{1}$ waves, showing that this coupled channel is
especially important. The mixing is driven by the tensor component of
the nucleon--nucleon interaction, which generates high-momentum
components in the correlated two-body wave function. These components
are crucial for describing short-range correlations and directly
affect the interference between one- and two-body currents in the
transverse response.

The solution of the BG equation can also be expressed in momentum
space, leading to correlated  partial waves
$\phi_{ll'}^{SJ}(p')$ that exhibit the high-momentum tails induced by
SRC.  These functions are obtained directly
as the Fourier transform of the coordinate-space solutions
$u_{ll'}^{SJ}(r)$.  We can separate the momentum-space solution into a
free part and a high momentum component
\begin{equation}
    {\phi}_{ll'}^{SJ}(p') \;=\; 
   \sqrt{\frac{\pi}{2}}\,\frac{1}{pp'}\,
   \delta(p-p')\,\delta_{ll'} 
+
    \Delta{\phi}_{ll'}^{SJ}(p') ,
    \label{eq:bg-momentum}
\end{equation}
where the  high-momentum component is given by
\begin{equation}
   \Delta{\phi}_{ll'}^{SJ}(p') \;=\;
   \sqrt{\frac{2}{\pi}}\,\frac{1}{pp'}\,
   \frac{Q(P,p')}{p^{2}-p'^{2}}
   \sum_{i}\hat{j}_{l}(p' r_{i})
   \sum_{l_{1}} (\lambda_{i})^{SJ}_{l_{1}l}\,
   u_{l_{1}l'}^{SJ}(r_{i}) .
\end{equation}

\begin{figure}[htb] 
  \centering
  \includegraphics[width=8cm,bb=18 470 470 770]{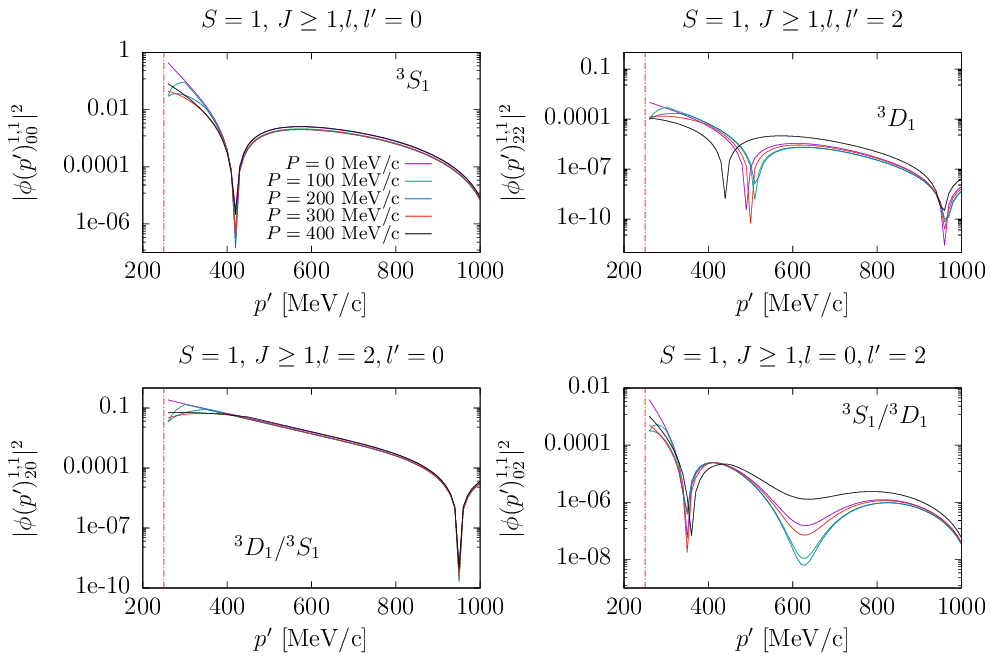}
  \caption{Example of momentum-space correlated wave functions 
  $|\tilde{\Phi}_{ll'}^{SJ}(p')|^{2}$ in a coupled channel. 
  The emergence of high-momentum components reflects the action of short-range correlations.}
  \label{fig:momentumwf}
\end{figure}

In Fig.~2 we show an example of the squared amplitude (momentum
distribution) of the high-momentum partial waves $\phi^{11}_{ll'}(p')$ for
the coupled $S=J=1$ channel of a nucleon pair with relative momentum
$p=225$ MeV/c, evaluated for several values of the total momentum. The
largest contributions arise from the $(l,l')=(0,0)$ ${}^{3}S_{1}$
component and from the $(l,l')=(2,0)$ ${}^{3}S_{1}$--${}^{3}D_{1}$
interference. This dominance reflects the strong tensor mixing in the
coupled ${}^{3}S_{1}$–${}^{3}D_{1}$ channel, which generates the
leading high-momentum components of the wave
function.

\section{SRC and MEC in QE Electron Scattering}

\begin{figure}
\centering
  \includegraphics[width=7cm,bb=120 460 495 700]{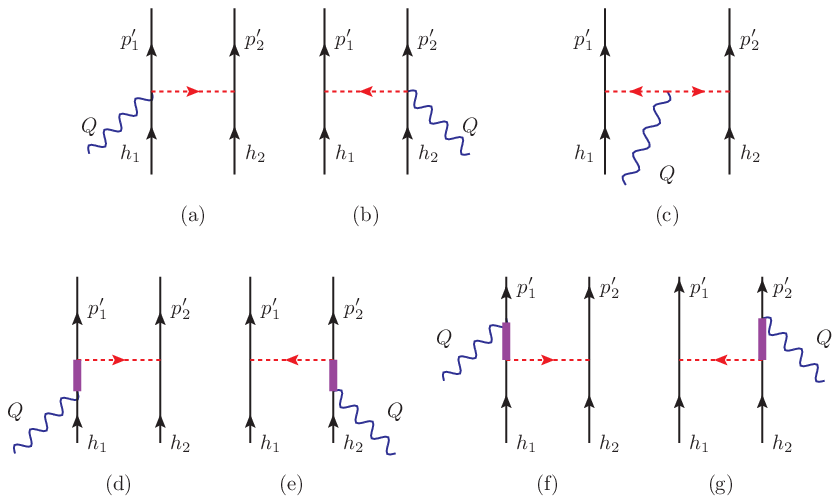}
\caption {Meson exchange currents contributing to the QE response:
seagull, pion-in-flight, and $\Delta$ excitation diagrams.}
\label{fig1}
\end{figure}

In QE electron scattering the longitudinal and
transverse responses correspond to 
$R_L(q,\omega)=W^{00}$ and $R_T(q,\omega)=W^{11}+W^{22}$, respectively,
where the hadronic tensor  in the RFG model is given by
\begin{equation}
    W^{\mu\mu}  = 
\sum_{ph}
|\langle ph^{-1}| J^\mu(\nq) | F \rangle |^2
\theta(p-k_F)\theta(k_F-h) 
\delta(E_p-E_h-\omega), \nonumber
\end{equation} 
Here we consider the non-relativistic current operator, written as the
sum of one-body and two-body meson-exchange current (MEC)
contributions at leading order, $\mathbf{J} = \mathbf{J}_{1b} +
\mathbf{J}_{2b}$.  The two-body operator  \cite{Ris89,Sch89} is the sum of the well-known
pion-exchange diagrams seagull, pionic or pion-in-flight and $\Delta$
currents of Fig. 3.

The $1p$--$1h$ matrix element of the current in the Fermi gas can be written as
\begin{equation}
  \langle ph^{-1}| J^\mu(\mathbf{q}) | F \rangle 
  =  \langle p| J_{1b}^\mu(\mathbf{q}) | h \rangle 
  + \sum_{k-k_F} \langle pk | J_{2b}^\mu(\mathbf{q}) | hk-kh \rangle.
\end{equation}
Taking the modulus squared, one obtains
\begin{align}
 | \langle ph^{-1}| J^\mu(\mathbf{q}) | F \rangle|^2 
  &=  |\langle p| J_{1b}^\mu(\mathbf{q}) | h \rangle|^2  \nonumber \\
  &\quad + 2\,{\rm Re}\!\left[ \langle p| J_{1b}^\mu(\mathbf{q}) | h \rangle^*
  \sum_{k} \langle pk | J_{2b}^\mu(\mathbf{q}) | hk-kh \rangle \right] \nonumber \\
  &\quad + \; \text{pure two-body MEC terms}.
\end{align}
The first term represents the pure one-body contribution, the second
term the interference between one- and two-body currents, and the last
term the pure MEC contribution, which is typically small and can be
neglected to first approximation. Thus the transverse response
function can be written schematically as
\begin{equation}
  R^T_{FG}(q,\omega)=R^T_{1b}(q,\omega)+R^T_{1b2b}(q,\omega)+\ldots
\end{equation}
In the Fermi gas, the interference term $R^T_{1b2b}$ is negative due to the
partial cancellation between seagull, pion-in-flight, and
$\Delta$  currents, with the {\em negative} $\Delta$ contribution dominating at
intermediate momentum transfers \cite{Cas25}.

When short-range correlations (SRC) are included within the
independent-pair approximation, the two-body matrix elements must be
evaluated with correlated two-particle states, $|\Phi_{hk}\rangle =
|hk\rangle + |\Delta \Phi_{hk}\rangle$, rather than the uncorrelated
states $|hk\rangle$. Accordingly, the matrix element of the two-body
current becomes
\begin{eqnarray}
     \langle ph^{-1}| J_{2b}^\mu(\mathbf{q}) | F \rangle 
     &=& \sum_{k} \langle pk | J_{2b}^\mu(\mathbf{q}) | \Phi_{hk}-\Phi_{kh} \rangle \nonumber \\
     &&\kern -2cm
     =     \sum_{k} \langle pk | J_{2b}^\mu(\mathbf{q}) | hk-kh \rangle 
     + \sum_{k} \langle pk | J_{2b}^\mu(\mathbf{q}) | \Delta \Phi_{hk}- \Delta \Phi_{kh} \rangle. 
\end{eqnarray}
Thus, in the interference term between one- and two-body currents, a
new contribution appears in which the MEC operator acts directly on
the high-momentum components of correlated pairs:
\begin{equation}
 2\, \mathrm{Re}\!\left[\langle p | J^\mu_{1b} | h \rangle^* 
   \sum_{k<k_F}
   \langle pk|J_{2b}^\mu|\Delta {\Phi}_{hk}-\Delta {\Phi}_{kh} \rangle \right].
\end{equation}
This new correlated MEC contribution can be represented schematically
by the diagram of Fig.~4, where the two-body operator couples to the
high-momentum components $(p_1,p_2)$ of the $(hk)$ pair.

The transverse response function can then be expressed as the sum of
the uncorrelated Fermi gas result (including MEC) plus an additional
term driven by SRC:
\begin{equation}
R^T(q,\omega) = R^T_{FG}(q,\omega) + \Delta R^T_{\rm SRC,MEC}(q,\omega),
\end{equation}
where $\Delta R^K_{\rm SRC,MEC}$ originates from the interference
between the one-body current and the MEC acting on the high-momentum
componets of correlated
pairs. This contribution  is essential
to account for the observed enhancement of the transverse response.

\begin{figure}
\centering
\includegraphics[width=6cm,bb=190 600 420 690]{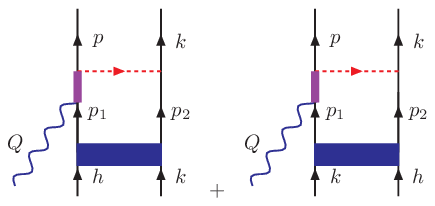}
\caption{High-momentum MEC contributions induced by SRC in the two-body current.}
\end{figure}

\section{Results}

\begin{figure}[h]
\centering
\includegraphics[width=4cm,angle=270]{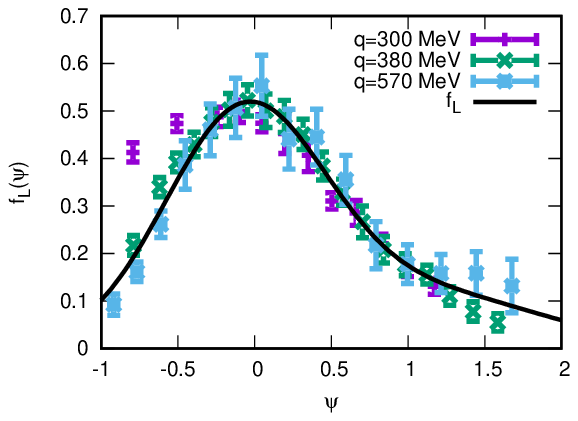}
\caption{Scaling function $f_L(\psi')$ extracted from $^{12}$C data
  and the SuSA fit.}
\label{fig:scaling}
\end{figure}

\begin{figure}[h]
  \centering
  \includegraphics[width=4cm,angle=270]{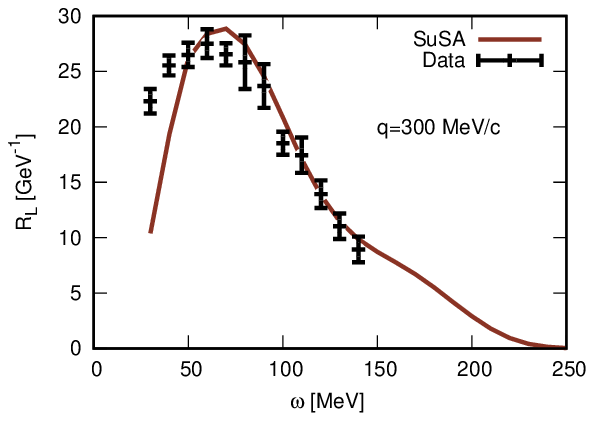}
\caption{Longitudinal response $R_L$ obtained using the SuSA scaling
  function. The data are well reproduced.}
\label{fig:RL}
\end{figure}

In order to compare our calculation with experimental data, it is well
known that the simple Fermi gas  model is inadequate. Therefore,
we adopt a superscaling approach \cite{Cas23a}, in which the longitudinal
scaling function $f_L(\psi')$ is extracted from inclusive electron
scattering data on $^{12}$C at momentum transfers up to $q \approx
570$~MeV/c \cite{Cas23}.  Figure~\ref{fig:scaling} shows the extracted scaling
function and the fit used in the present study. The corresponding
longitudinal response $R_L$ is well reproduced by this fit, as shown
in Fig.~\ref{fig:RL}.

\begin{figure}[h]
  \centering
  \includegraphics[width=4cm,angle=270]{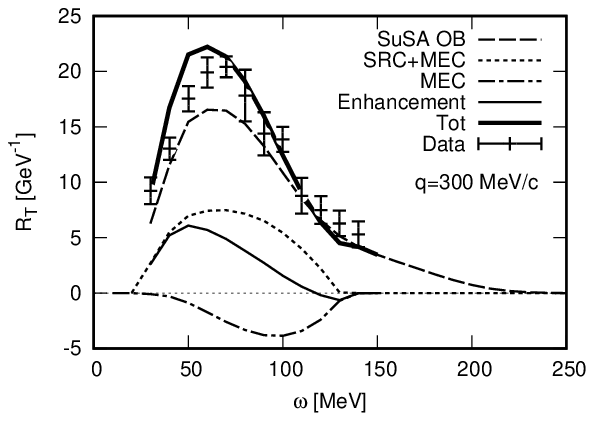}
\caption{Transverse response $R_T$ showing separate contributions: OB
  only, OB+MEC in the Fermi gas, and OB+MEC including SRC
  (high-momentum components). The enhancement arises from the
  interference of the coupled $^3S_1$-$^3D_1$ channels, i.e., the
  tensor force.}
\label{fig:RT}
\end{figure}

Assuming scaling of the zero kind, $f_T=f_L$, we investigate the
transverse response. Figure~\ref{fig:RT} shows that the naive Fermi
gas calculation with MEC produces a reduction of the transverse response,
failing to reproduce the data. By contrast, the inclusion of
short-range correlations in the MEC matrix elements, following our
independent pair approximation and the Bethe--Goldstone solution, generates
an enhancement that cancels the reduction from the FG MEC and
successfully reproduces the experimental transverse response.

The enhancement is mainly produced by the interference of the coupled
$^3S_1$-$^3D_1$ partial waves, highlighting the crucial role of the tensor
force in the correlated wave function. Remarkably, this effect emerges
directly from the NN potential (Granada 2013), fitted to NN cross
section data, without any additional parameter or fit for the enhancement.
This result aligns with Fabrocini's correlated basis function (CBF)
calculations in nuclear matter, showing that despite the different
approaches to SRC, both methods produce a similar tendency towards
transverse enhancement. Our calculation therefore provides an
independent validation of the role of correlations and MEC in
enhancing the $1p1h$ transverse response in electron scattering.

The formalism developed here can be directly extended to quasielastic 
neutrino-nucleus scattering. In the Fermi gas, meson-exchange currents 
also produce a reduction of the $1p1h$ response, as in the electromagnetic 
case \cite{Cas25b}. The main difference is that the weak response receives contributions 
from both vector and axial currents, in the one-body (OB) and two-body 
(MEC) sectors. Schematically, the modification of the matrix elements can 
be written as
\begin{equation}
\langle ph^{-1} | J^\mu_{\rm weak} | F \rangle 
= \langle p | J^\mu_{1b,V}+J^\mu_{1b,A} | h \rangle
+ \sum_{k<k_F} \langle pk | J^\mu_{2b,V}+J^\mu_{2b,A} | \phi_{hk}-\phi_{kh} \rangle ,
\nonumber
\end{equation}
where $V$ and $A$ denote vector and axial contributions, respectively, 
and the correlated wave functions $\phi_{hk}$ are obtained from the 
Bethe-Goldstone equation as before. Preliminary calculations indicate 
that the enhancement observed in the transverse electron response persists 
when the axial current is included, although a detailed study is required 
to quantify the effect. These results will be presented in a forthcoming publication.

\section{Conclusions}

We have developed a correlated model for the $1p1h$ transverse response 
including MEC and SRC within the independent pair approximation. Solving 
the Bethe-Goldstone equation with the Granada 2013 NN potential, we find 
that high-momentum components generate an enhancement that compensates 
the reduction predicted by the Fermi gas and reproduces the experimental 
data. The effect arises from coupled-channel waves, notably ${}^3S_1$--${}^3D_1$, 
highlighting the tensor force, and emerges directly from the realistic NN 
interaction without fitting. Our results are consistent with CBF 
calculations and suggest that a similar enhancement may occur in neutrino 
scattering.

\section*{Acknowledgements}

The work was supported by Grant No.  PID2023-147072NB-I00 funded by
MICIU/AEI /10.13039/501100011033 and by ERDF/EU; by Grant No.  FQM-225
funded by Junta de Andalucia.

\end{document}